\documentclass{pasj00}
\Received{$\langle$2011 July 25$\rangle$}
\Accepted{$\langle$2011 Seutember 6$\rangle$}
\Published{$\langle$publication date$\rangle$}
\SetRunningHead{Astronomical Society of Japan}{Usage of \texttt{pasj00.cls}}
\def\Hline{\noalign{\hrule height 1.5pt}}
\usepackage{times}
\usepackage{lineno}
\usepackage{color}

\begin{document}
\SetRunningHead{Nakahira et al.}{Spectral Study of XTE\ J1752--223}

\title{A Spectral Study of the Black Hole Candidate XTE J1752-223
in the High/Soft State with MAXI, Suzaku and Swift}

\author{
Satoshi \textsc{Nakahira},\altaffilmark{1}
Shu \textsc{Koyama},\altaffilmark{2}
Yoshihiro \textsc{Ueda},\altaffilmark{3}
Kazutaka \textsc{Yamaoka},\altaffilmark{4}
Mutsumi \textsc{Sugizaki},\altaffilmark{1}
Tatehiro \textsc{Mihara},\altaffilmark{1}
Masaru \textsc{Matsuoka},\altaffilmark{1}
Atsumasa \textsc{Yoshida},\altaffilmark{4}
Kazuo \textsc{Makishima},\altaffilmark{1,5}
Ken \textsc{Ebisawa},\altaffilmark{6}
Aya \textsc{Kubota},\altaffilmark{7}
Shin'ya \textsc{Yamada},\altaffilmark{8}
Hitoshi \textsc{Negoro},\altaffilmark{9}
Kazuo \textsc{Hiroi},\altaffilmark{3}
Masaki \textsc{Ishikawa},\altaffilmark{10}
Nobuyuki \textsc{Kawai},\altaffilmark{11}
Masashi \textsc{Kimura},\altaffilmark{12}
Hiroki \textsc{Kitayama},\altaffilmark{12}
Mitsuhiro \textsc{Kohama},\altaffilmark{13}
Takanori \textsc{Matsumura},\altaffilmark{14}
Mikio \textsc{Morii},\altaffilmark{11}
Motoki \textsc{Nakajima},\altaffilmark{15}
Motoko \textsc{Serino},\altaffilmark{1}
Megumi \textsc{Shidatsu},\altaffilmark{3}
Tetsuya \textsc{Sootome},\altaffilmark{1,7}
Kousuke \textsc{Sugimori},\altaffilmark{11}
Fumitoshi \textsc{Suwa},\altaffilmark{9}
Hiroshi \textsc{Tomida},\altaffilmark{13}
Yoko \textsc{Tsuboi},\altaffilmark{14}
Hiroshi \textsc{Tsunemi},\altaffilmark{12}
Shiro \textsc{Ueno},\altaffilmark{13}
Ryuichi \textsc{Usui},\altaffilmark{11}
Takayuki \textsc{Yamamoto},\altaffilmark{1,9}
Kyohei \textsc{Yamazaki},\altaffilmark{14}
Makoto S. \textsc{Tashiro},\altaffilmark{2}
Yukikatsu \textsc{Terada},\altaffilmark{2} and
Hiromi \textsc{Seta}\altaffilmark{2}
	}
\altaffiltext{1}{MAXI team, Institute of Physical and Chemical Research (RIKEN), 2-1 Hirosawa, Wako, Saitama 351-0198}
\altaffiltext{2}{Department of Physics, Saitama University, 255, Shimo-Okubo, Sakura-ku, Saitama 338-8570}
\altaffiltext{3}{Department of Astronomy, Kyoto University, Oiwake-cho, Sakyo-ku, Kyoto 606-8502}
\altaffiltext{4}{Department of Physics and Mathematics, Aoyama Gakuin University,\\ 5-10-1 Fuchinobe, Chuo-ku, Sagamihara, Kanagawa 252-5258}
\altaffiltext{5}{Department of Physics, The University of Tokyo, 7-3-1, Hongo, Bunkyo-ku, Tokyo, 113-0033}
\altaffiltext{6}{Department of Space Science Information Analysis, Institute of Space and Astronautical Science, Japan Aerospace Exploration Agency , 3-1-1 Yoshino-dai, Chuo-ku, Sagamihara, Kanagawa 252-5210, Japan}
\altaffiltext{7}{Department of Electronic Information Systems, Shibaura Institute of Technology, 307 Fukasaku, Minuma-ku, Saitama, Saitama 337-8570}
\altaffiltext{8}{High Energy Astrophysics Laboratory, Institute of Physical and Chemical Research (RIKEN)\\ 2-1 Hirosawa, Wako, Saitama 351-0198}
\altaffiltext{9}{Department of Physics, Nihon University, 1-8-14, Kanda-Surugadai, Chiyoda-ku, Tokyo 101-8308}
\altaffiltext{10}{School of Physical Science, Space and Astronautical Science, The graduate University for Advanced Studies (Sokendai), Yoshinodai 3-1-1, Chuo-ku, Sagamihara, Kanagawa 252-5210}
\altaffiltext{11}{Department of Physics, Tokyo Institute of Technology, 2-12-1 Ookayama, Meguro-ku, Tokyo 152-8551}
\altaffiltext{12}{Department of Earth and Space Science, Osaka University, 1-1 Machikaneyama, Toyonaka, Osaka 560-0043}
\altaffiltext{13}{ISS Science Project Office, Institute of Space and Astronautical Science (ISAS), Japan Aerospace Exploration Agency (JAXA), 2-1-1 Sengen, Tsukuba, Ibaraki 305-8505}
\altaffiltext{14}{Department of Physics, Chuo University, 1-13-27 Kasuga, Bunkyo-ku, Tokyo 112-8551}
\altaffiltext{15}{School of Dentistry at Matsudo, Nihon University, 2-870-1 Sakaecho-nishi, Matsudo, Chiba 101-8308}


%



\maketitle

\begin{abstract}
We report on the X-ray spectral analysis of the black hole candidate
XTE\ J1752--223 in the 2009--2010 outburst, utilizing data obtained
with the MAXI/Gas Slit Camera (GSC), the Swift/XRT, and Suzaku, which
work complementarily. As already reported by \citet{1752nkhr} MAXI
monitored the source continuously throughout the entire outburst for
about eight months. All the MAXI/GSC energy spectra in the high/soft
state lasting for 2 months are well represented by a multi-color disk
plus power-law model.  The innermost disk temperature changed from
$\sim$0.7 keV to $\sim$0.4 keV and the disk flux decreased by an order
of magnitude.  Nevertheless, the innermost radius is constant at
$\sim$41 $D_{3.5}(\cos{\it i})^{-\frac{1}{2}}$ km, where $D_{3.5}$ is
the source distance in units of 3.5 kpc and $i$ the inclination.  
The multi-color disk parameters obtained with the MAXI/GSC are consistent
with those with the Swift/XRT and Suzaku. The Suzaku data also
suggests a possibility that the disk emission is slightly Comptonized,  
which could account for broad iron-K features reported previously. Assuming
that the obtained innermost radius represents the innermost stable
circular orbit for a non-rotating black hole, we estimate the mass of
the black hole to be 5.51$\pm$0.28 $M_{\odot}$ $D_{3.5}(\cos{\it
i})^{-\frac{1}{2}}$, where the correction for the stress-free inner
boundary condition and color hardening factor of 1.7 are taken into
account. If the inclination is less than 49$^{\circ}$ as suggested from
the radio monitoring of transient jets and the soft-to-hard transition
in 2010 April occurred at 1--4\% of Eddignton luminosity, the fitting
of the Suzaku spectra with a relativistic accretion-disk model derives
constraints on the mass and the distance to be 3.1--55 $M_{\odot}$
and 2.3--22 {\rm kpc}, respectively. This confirms that the
compact object in XTE J1752--223 is a black hole.
\end {abstract}
\clearpage
\section{Introduction}

Galactic black hole candidates (BHCs) show various states
characterized by their spectral shapes, temporal
properties, and luminosities. They mostly take two major states,
the ``low/hard state'' and the ``high/soft
state'', which are referred to as ``the hard X-ray state'' and ``the
thermal dominant state'', respectively, in more recent classification
(see \cite{review} and references therein) since the state of a BHC is not
always determined by an X-ray luminosity alone. In the low/hard
state,
the X-ray energy spectra are dominated by a power-law component
with a photon index of $\sim$1.7 and a high energy cutoff 
at $\sim$100 keV that shows strong short-time variability
\citep{lhsgrove}. They can be explained by thermal Comptonization of
soft photons from the accretion disk by hot plasmas with a temperature
of $\sim$10$^{10}$ K (e.g., \cite{cygx1_makishima}, \cite{1655_takahashi},
\cite{339_shidatsusz}). 
In a typical outburst of a transient BHC, it
generally exhibits a spectral transition from the
low/hard to the high/soft state through the intermediate (or very
high) state when the luminosity reaches $\sim$10\% of the Eddingtion limit. 
The X-ray spectra in the high/soft state are
characterized by a ultra-soft component which is considered
to originate from an optically-thick and geometrically-thin accretion
disk (so called ``standard disk''; \cite{stddisk}). This
emission can be successfully described by the multi-color disk (MCD)
model (\cite{mcdmodel}, \cite{rinconst}) with the innermost
temperature of $\sim$1 keV. An important signature in the high/soft
state is that the innermost radius ($r_{\rm in}$) of the accretion
disk is kept constant, independent of the flux or innermost disk
temperature (e.g., \cite{xraynovae}). Hence, it is believed that the
$r_{\rm in}$ reflects the innermost stable circular orbit (ISCO),
which is determined through general relativity by the mass and angular momentum of the black
hole; the radius of ISCO for a non-spinning black hole is 6$R_{\rm
g}$ ($R_{\rm g}\equiv GM/c^2$ is the gravitational radius, where $G$,
$M$, $c$ are the gravitational constant, black hole mass, and light
velocity, respectively).
By assuming that the obtained innermost $r_{\rm in}$ to the ISCO, the
black hole mass can be estimated from the X-ray spectrum alone for a given spin
parameter.

XTE\ J1752--223 was first discovered with the Proportional Counter Array (PCA)
onboard Rossi X-ray Timing Explorer (RXTE) on 2009 October 23 (MJD
55127) during a Galactic bulge scan observation (\cite{1752discov}).
The Gas Slit Camera (GSC) onboard Monitor of All-sky X-ray Image (MAXI; \cite{maxi})
detected the source simultaneously \citep{atel_maxinkhr}. 
As described in \authorcite{1752nkhr} (2010; hereafter Paper I), MAXI continuously 
monitored the source during the entire outburst until 2010
June. The MAXI light curves and hardness-intensity diagram revealed that
the source initially stayed in the low/hard states with two stable flux
levels for about three months, and then moved into the high/soft state
(Paper I). A radio flare was detected simultaneously with the
spectral transition (\cite{atel_maxinegoro}, \cite{atel_jet_trans}), and
the proper motion of the jet ejecta was later found with VLBI
observations (\cite{radio_evnvlba}). 
\citet{accurate_pos_vlbi} identified it as an approaching jet ejected 
coincidently with the X-ray state transition. From the observed flux of the 
approaching jet and its upper limit of the receding one, 
they constrained the jet speed and the inclination angle from 
the line-of-sight to be $>0.66c$ and $< 49^\circ$, respectively, 
by assuming that the axes of the twin jets are aligned.
\citet{1752shapos} estimated the black hole 
mass and distance to be 9.6$\pm$0.9 $M_{\odot}$ and 3.5$\pm$0.4 {\rm kpc}, 
respectively, using the spectral-timing correlation technique (\cite{shapos_corr}).
However, the compact object mass has not been estimated via optical 
mass-function technique.
To firmly establish the nature of XTE\ J1752--223, however, it is quite
important to constrain the black hole mass by an independent approach.

In this paper, we present the results from a detailed spectral
analysis of XTE\ J1752--223 utilizing the MAXI/GSC data, together with
those of Swift/XRT and Suzaku, mainly focusing on the spectra in the
high/soft state. We then discuss the constraints on the black hole mass and
distance based on our results. In the appendix, we present the current
status of spectral calibration of the MAXI/GSC using the Crab Nebula,
which is fully reflected in this paper and in a similar work for the
black hole candidate GX 339--4 reported by \citet{339_shidatsumx}.
The spectral fitting was carried out on XSPEC version 12.6. Errors
are quoted at statistical 90\% confidence limits for a single
parameter throughout the paper.

\section{MAXI observation and analysis}

\subsection{MAXI observations}

As the first astronomical mission on the International Space Station (ISS), 
MAXI has been operating since 2009 August. The payload
carries two kinds of X-ray all-sky monitors; the Gas-Slit Camera (GSC;
\cite{gscmihara}) and the Solid-state Slit Camera (SSC;  \cite{ssctsumemi}, \cite{ssctomida}).
The GSC consists of Xe-gas proportional counters for the X-ray
detector and slat collimators with a slit to localize the direction of
the incident X-rays. The counters employ resistive-carbon wires for
detector anodes to determine the X-ray position along the anode
wire. Twelve identical units (refereed to as GSC\_0, ..., GSC\_9, 
GSC\_A and GSC\_B) are assembled so that they instantaneously cover two
rectangular field of views (FoVs) of 3$\times$160 deg$^2$ with an equal area.

After the operation started in 2009 August, two counters, GSC\_6 and
GSC\_9 were stopped on 2009 September 8 and 14, respectively, for 
high-voltage breakdown. Other two counters on GSC\_A and GSC\_B were
stopped temporarily on September 23, because diagnostic data indicated 
that they may also break down rather soon.
Consequently, the outburst of XTE J1752--223 from 2009 October to 2010 
June was covered by eight GSC camera units out of the twelve.
 
The GSC FoV is limited by slats collimators.  The visibility for a
target at a given sky position changes according to the ISS orbital
motion; hereafter we call the visibility time as ``transit''.
Each transit lasts for 40-150 s, and the effective area of each GSC counter 
change due to the triangular-shaped collimator transmission function, 
with a peak value of 4-5 cm$^2$.
The MAXI/GSC scanned the direction of
XTE\ J1752--223 2041 times in total from the discovery on 2009 October
23 (MJD 55127) to 2010 June 3 (MJD 55350). The total exposure times
effective area becomes 534.8 cm$^2$ ksec.

\subsection{Analysis}

\begin{figure}[!ht]
  \begin{center}
    \FigureFile(80mm,80mm){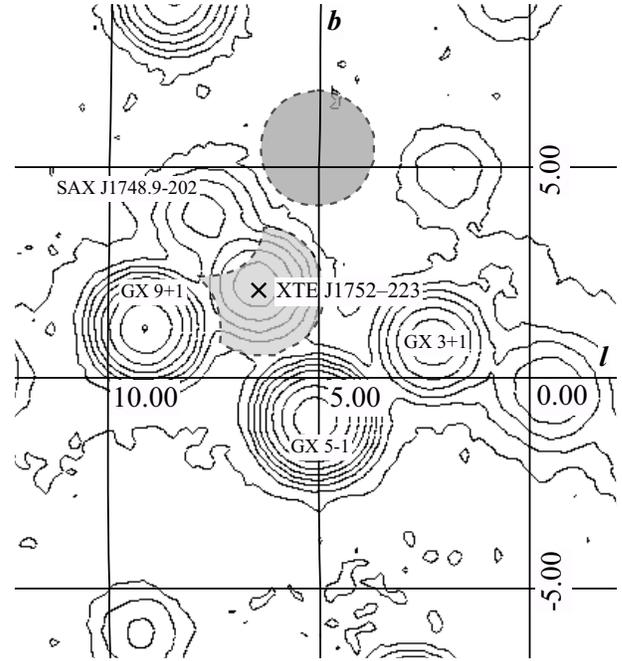}
  \end{center}
  \caption{A 2--20 keV MAXI/GSC image in the galactic coordinate around XTE\ J1752--223. 
The data are accumulated from MJD 55200 to MJD 55299.
The intensity contours are shown by a log scale.
 The source and background regions for the analysis are indicated by the shadowed regions.}
  \label{fig:contour_region}
\end{figure}

For the data analysis, we used the MAXI specific analysis tools, which
were developed by the MAXI team. We analyzed the GSC event data
version 0.3b, which include the data taken by counters operated at the
nominal high voltage (=1650 V) but excluding those of anode \#1 and
\#2 whose energy responses have not been enough calibrated yet.  We
discarded events taken while the GSC FoVs were interfered by the solar
panels and other ISS payloads. The events detected at the anode-end
area were also screened since the background is higher therein. 
These event were cut with a condition that the photon incident
angle ($\phi_{\rm col}$; see \cite{gscmihara} for the definition) is higher
than 36$^{\circ}$. As shown in Figure \ref{fig:contour_region} we carefully 
extracted events for source and background from circular regions with a 
1.5$^{\circ}$ radius, excluding regions within 1.7$^{\circ}$ of nearby sources, 
GX 5--1, GX 9+1 and SAX J1748.9--2021 in NGC 6440.
We used only those data when the source and background 
regions were both fully scanned in a transit.

We performed spectral analysis of the data taken between MJD=55200 and
MJD=55293, during which both the ``hard-to-soft'' and ``soft-to-hard''
transitions took place (Paper I). The net effective exposure was 194.5 cm$^2$
ksec for 752 transits. We divided the whole dataset into groups with
typical lengths of a few days so that the spectrum in each group has
enough photon statistics, except for the epoch around the
``hard-to-soft'' state transition when the spectrum changed
rapidly. Eventually, data were separated into 49 groups whose
exposures times effective area were 0.2$\sim$12 cm$^2$ ksec. The energy
response matrix for each group was calculated by the response builder
(\cite{gscsugizaki}). The validity of the energy response was verified
using the Crab Nebula (see Appendix). 

\subsection{Light Curves during the 2009--2010 Outburst}

\begin{figure}[htb]
\begin{center}
\FigureFile(80mm,80mm){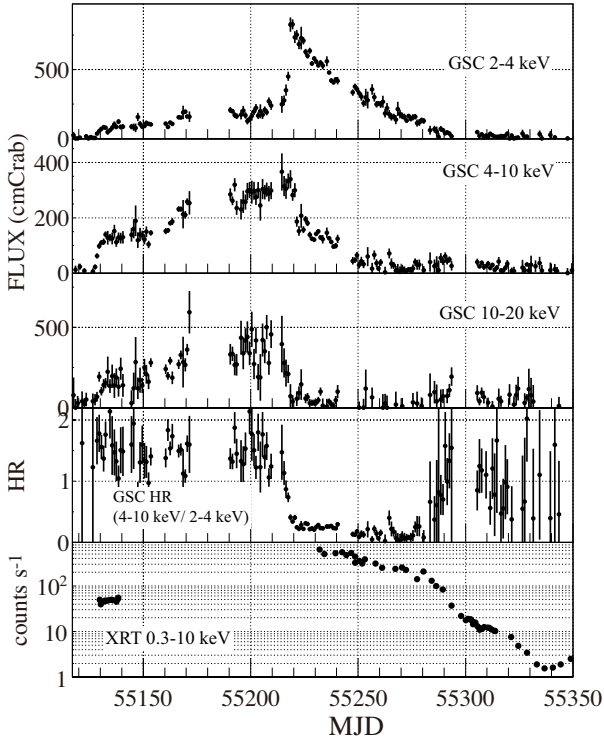}
\end{center}
  \caption{MAXI/GSC and Swift/XRT light curves of XTE\ J1752--223 in
   the 2009-2010 outburst.  Four panels from the top show the GSC
   light curves in three energy bands of 2--4, 4--10 and 10--20 keV,
   and the hardness ratio between the 4--10 keV and 2--4 keV bands.
   The bottom shows the Swift/XRT light curve in the 0.3--10 keV
   band.}
\label{fig:lightcurves}
\end{figure}

Figure~\ref{fig:lightcurves} shows the MAXI/GSC light curves of XTE\
J1752--223 during the 2009--2010 outburst in the 2--4 keV, 4--10 keV,
and 10--20 keV bands, together with the hardness ratio between the
4--10 keV and 2--4 keV bands. They are updated from those presented in
Paper I, after applying the latest calibration and the same data
screening as used for the spectral analysis (section~\ref{subsec-spec}
and Appendix). The MAXI/GSC first detected XTE\ J1752--223 at 15:05
(UT) on 2009 October 23 (MJD 55127) when the 2--20 keV X-ray intensity was 30 mCrab. 
It monitored the source almost uniformly
except for the time when the FoV was close to the sun (MJD
55154--55159 and MJD 55172--55188) and when poles of the scanning axis was
close to the source (around MJD 55300). The light curves reveal that
the source stayed in the initial low/hard state for the extraordinary
long period of about 3 months, and took two flux-plateau phases meantime (Paper I).

The ``hard-to-soft'' state transition occurred on MJD 55218 and the
following behavior agrees well with those of the typical BHC outbursts
(\cite{review2}). The X-ray intensity was peaked at 420 mCrab on 2010
January 22 (MJD 55218), then decayed exponentially with an e-folding
time of 34 days through the high/soft state. The ``soft-to-hard''
transition started on 2010 March 30 (MJD 55285), and then the source
returned to the low/hard state on 2010 April 6 (MJD 55292). 
On 2010 June 28 (MJD 55375), the source flux fell below the MAXI/GSC 
detection sensitivity per day (20 mCrab). 
Thus, the total duration of the outburst became about eight months.

\subsection{Spectral Analysis}
\label{subsec-spec}

For the spectral analysis, we employed the standard model for BHCs in
the high/soft state, a multi-color disk (MCD; {\em diskbb} in XSPEC)
model plus a power-law representing the hard tail.
The {\em wabs} (\cite{wabs1}) model with 
solar abundances by \citet{wabs2} was applied for the interstellar absorption.
The hydrogen column density ($N_{\rm H}$) toward XTE\ J1752--223 was
fixed at 0.6$\times$10$^{22}$ cm$^{-2}$, based on the Swift and Suzaku 
results as described below.
The model has four parameters: the innermost temperature $T_{\rm in}$ and
innermost radius $r_{\rm in}$\footnote{$r_{\rm in}$=$\sqrt{N_{\rm
diskbb}}$ $(D/10 {\rm kpc}) (\cos{\it i})^{-\frac{1}{2}}$, where
$N_{\rm diskbb}$ is the normalization of the {\em diskbb} model. The
distance to the source of $D=3.5$ kpc and inclination angle $i$=0 are
assumed.} for the MCD component, with the photon index $\Gamma$ and  
normalization at 1 keV for the power-law component. We first applied the model to
all the data. When the MCD component was found to be not necessary
(i.e., the 90\% confidence range of the MCD normalization contains
zero), only the power-law model was applied with its $\Gamma$ set free.
When the MCD component was required, we fixed $\Gamma=2.2$ for all the
spectra because it is often difficult to determine from individual spectrum
due to poor statistics and the limited energy coverage of the
MAXI/GSC. This photon index corresponds to a typical value obtained
when the power-law component is relatively strong ($>$50\% of the
total flux in the 2--20 keV band).
Although \citet{1752shapos} suggest from the RXTE/PCA data that the power-law slope is variable
between $\Gamma \approx 1.6-2.2$ by using a spectral model different from ours,
we confirmed that varying $\Gamma$ value within this range
only changes the best-fit MCD normalization by $< 3\%$, and hence does not
affect our conclusions.

\begin{figure*}[htb!]
\begin{center}
\FigureFile(170mm,170mm){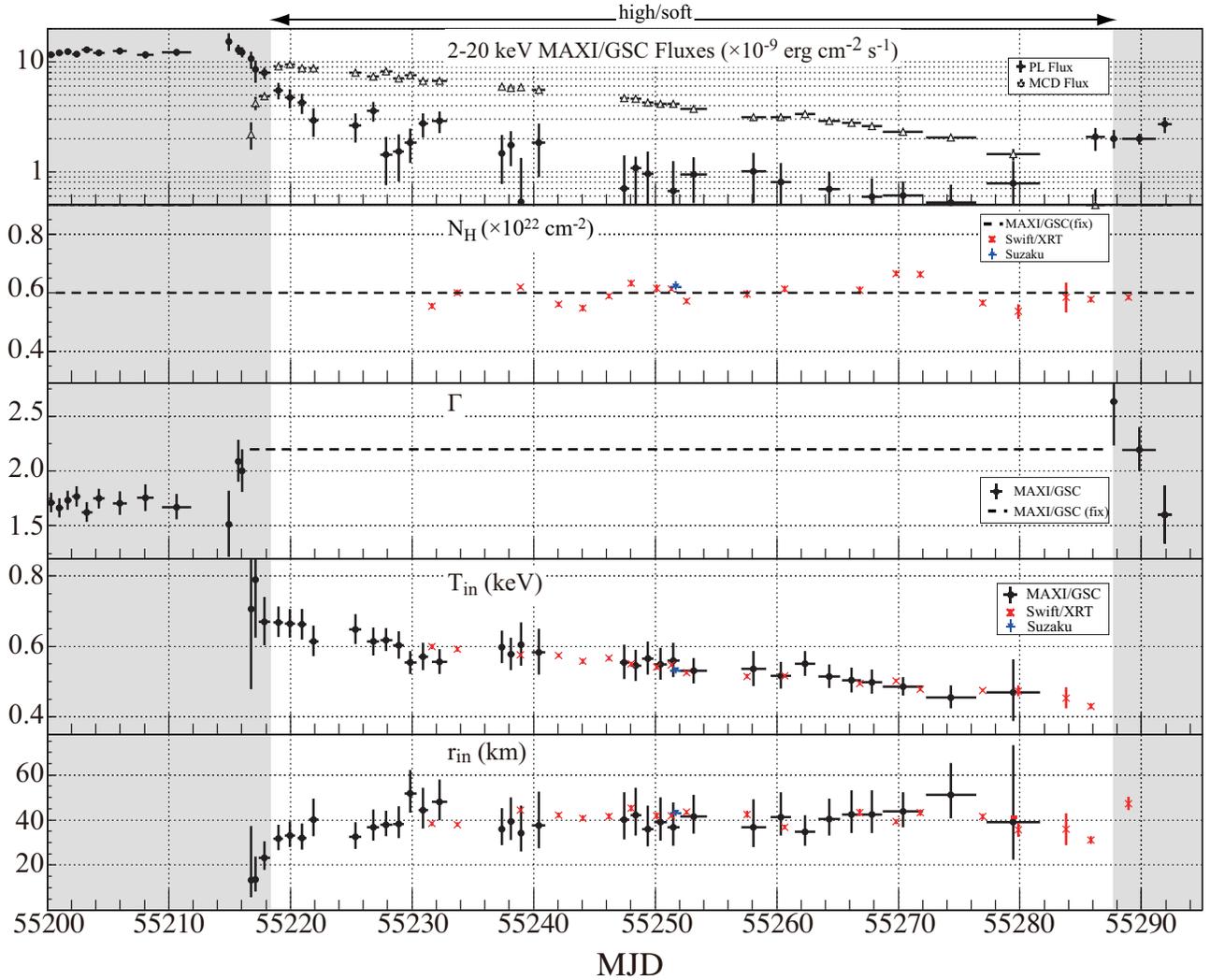}
\end{center}
\caption{Evolution of the spectral parameters of XTE\ J1752--223.
The MAXI/GSC, Swift/XRT, and Suzaku results using the MCD plus power-law model
are shown by black, red, and blue points, respectively. 
From top to bottom panels, 2--20 keV fluxes of MCD and power-law components,
the hydrogen column density $N_{\rm H}$, the photon index $\Gamma$,
the inner most temperature $T_{\rm in}$, and the innermost radius $r_{\rm in}$ are indicated.}

\label{fig:time_vs_parameters}
\end{figure*}

\begin{figure*}[!htb]
\begin{center}
\FigureFile(170mm,170mm){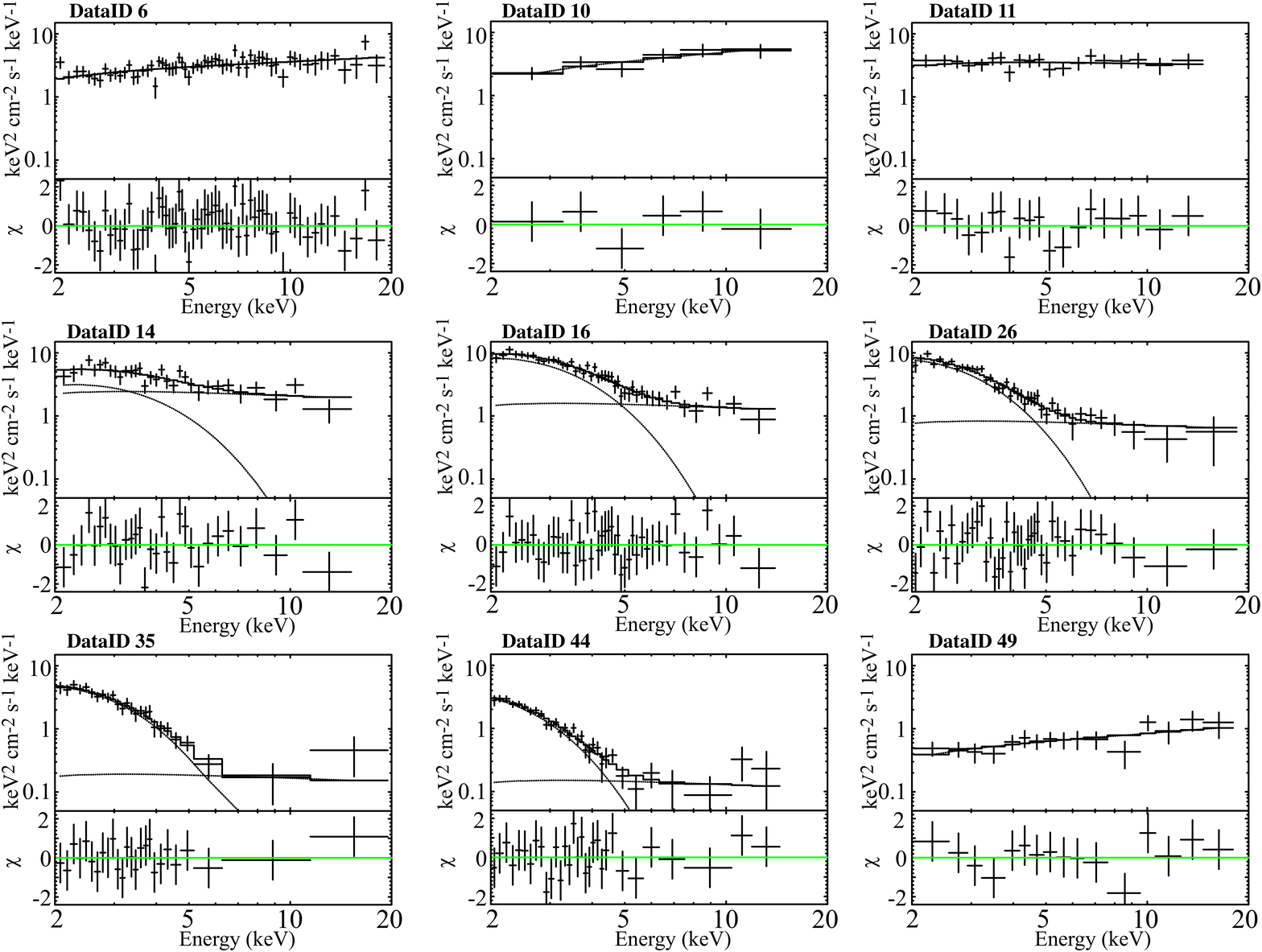}
\end{center}
\caption{
Examples of the $\nu$F$\nu$ spectrum of XTE\ J1752--223 observed with
the MAXI/GSC together with the best-fit model. Panels correspond to 
DataIDs 6, 10, 11, 14, 16, 26, 35, 44, and 49. Data ID 35 was taken
approximately at the same time as the Suzaku observation. }

\label{fig:nufnuspec}
\end{figure*}
\begin{longtable}{rcccccccr}
  \caption{Best fit parameters of MAXI observations.}
  \label{tab:parametertab}
  \hline\hline
DataID & MJD & exposure & $\Gamma$ & power-law & $T_{\rm in}$  & $r_{\rm in}$  & Disk  & $\chi^{2}_{\nu}$/dof \\
   {\ } & start--end   &{(cm$^2$ ksec)} &  {\ } & flux$^*$ & (keV) & (km)  & flux$^*$ & {\ }  \\    
\hline
\endhead
1 & 55200.03--55200.66 & 3.42 & 1.71$\pm$0.09 & 11.6$\pm$0.7 & -- & -- & -- & 1.33(59) \\
2 & 55200.73--55201.30 & 3.41 & 1.67$\pm$0.09 & 12.1$\pm$0.7 & -- & -- & -- & 1.36(61) \\
3 & 55201.42--55202.00 & 3.42 & 1.73$\pm$0.09 & 12.4$\pm$0.7 & -- & -- & -- & 0.92(64) \\
4 & 55202.06--55202.76 & 3.41 & 1.77$\pm$0.09 & 11.8$\pm$0.7 & -- & -- & -- & 0.79(60) \\
5 & 55202.82--55203.71 & 3.41 & 1.63$\pm$0.09 & 13.0$\pm$0.7 & -- & -- & -- & 0.81(65) \\
6 & 55203.77--55204.72 & 3.40 & 1.75$\pm$0.09 & 12.1$\pm$0.7 & -- & -- & -- & 0.93(62) \\
7 & 55205.42--55206.56 & 2.37 & 1.70$\pm$0.11 & 12.6$\pm$0.8 & -- & -- & -- & 1.10(47) \\
8 & 55207.38--55208.73 & 2.01 & 1.76$_{-0.12}^{+0.13}$ & 11.6$\pm$0.9 & -- & -- & -- & 0.57(36) \\
9 & 55209.41--55211.84 & 2.01 & 1.67$_{-0.11}^{+0.12}$ & 12.2$\pm$0.9 & -- & -- & -- & 1.13(39) \\
10 & 55214.88--55214.94 & 0.24 & 1.51$_{-0.30}^{+0.31}$ & 15.4$_{-2.8}^{+2.9}$ & -- & -- & -- & 0.67(4) \\
11 & 55215.64--55215.83 & 0.83 & 2.09$\pm$0.19 & 13.0$\pm$1.3 & -- & -- & -- & 0.55(18) \\
12 & 55215.89--55216.15 & 0.83 & 2.00$_{-0.19}^{+0.20}$ & 12.3$_{-1.3}^{+1.4}$ & -- & -- & -- & 0.87(17) \\
13 & 55216.65--55216.91 & 1.06 & 2.20(fix) & 10.7$_{-2.0}^{+1.7}$ & 0.71$_{-0.23}^{+0.29}$ & 13.3$_{-7.6}^{+23.9}$ & 2.2$\pm$0.6 & 1.09(24) \\
14 & 55216.97--55217.29 & 1.28 & 2.20(fix) & 8.5$_{-2.1}^{+1.8}$ & 0.79$_{-0.16}^{+0.19}$ & 13.5$_{-5.2}^{+10.4}$ & 4.2$\pm$0.6 & 0.94(29) \\
15 & 55217.42--55218.37 & 3.39 & 2.20(fix) & 8.0$_{-1.0}^{+0.9}$ & 0.67$\pm$0.07 & 23.2$_{-5.3}^{+7.4}$ & 4.9$\pm$0.4 & 0.86(51) \\
16 & 55218.62--55219.38 & 2.93 & 2.20(fix) & 5.5$\pm$1.0 & 0.67$_{-0.04}^{0.05}$ & 31.8$_{-5.0}^{+6.3}$ & 9.2$\pm$0.5 & 0.73(42) \\
17 & 55219.57--55220.39 & 3.22 & 2.20(fix) & 4.7$\pm$0.9 & 0.66$\pm$0.04 & 33.1$_{-5.0}^{+6.2}$ & 9.6$_{-0.4}^{+0.5}$ & 0.77(43) \\
18 & 55220.58--55221.35 & 3.06 & 2.20(fix) & 4.3$\pm$0.9 & 0.66$\pm$0.04 & 32.0$_{-5.1}^{+6.4}$ & 8.7$\pm$0.4 & 0.63(40) \\
19 & 55221.54--55222.30 & 2.27 & 2.20(fix) & 3.0$_{-0.9}^{+0.8}$ & 0.61$\pm$0.04 & 40.1$_{-7.2}^{+9.3}$ & 8.8$\pm$0.5 & 1.11(33) \\
20 & 55224.84--55225.85 & 3.03 & 2.20(fix) & 2.6$\pm$0.8 & 0.65$\pm$0.04 & 32.5$_{-5.2}^{+6.6}$ & 8.0$\pm$0.4 & 1.15(35) \\
21 & 55226.30--55227.31 & 3.31 & 2.20(fix) & 3.6$\pm$0.7 & 0.61$\pm$0.04 & 36.9$_{-6.2}^{+7.7}$ & 7.4$\pm$0.4 & 0.63(39) \\
22 & 55227.37--55228.39 & 3.53 & 2.20(fix) & 1.4$\pm$0.7 & 0.62$\pm$0.03 & 38.0$_{-5.2}^{+6.2}$ & 8.2$\pm$0.4 & 1.13(33) \\
23 & 55228.45--55229.34 & 3.33 & 2.20(fix) & 1.5$\pm$0.7 & 0.60$\pm$0.04 & 38.2$_{-6.2}^{+7.9}$ & 7.1$\pm$0.4 & 1.19(30) \\
24 & 55229.41--55230.36 & 3.82 & 2.20(fix) & 1.8$\pm$0.6 & 0.55$\pm$0.03 & 51.7$_{-8.4}^{+10.5}$ & 7.6$\pm$0.3 & 1.25(32) \\
25 & 55230.55--55231.31 & 3.40 & 2.20(fix) & 2.7$\pm$0.7 & 0.57$\pm$0.04 & 44.3$_{-8.0}^{+10.1}$ & 6.7$\pm$0.4 & 1.29(34) \\
26 & 55231.63--55232.90 & 4.30 & 2.20(fix) & 2.9$_{-0.7}^{+0.6}$ & 0.56$_{-0.03}^{+0.04}$ & 48.0$_{-8.1}^{+10.1}$ & 6.8$\pm$0.3 & 0.93(38) \\
27 & 55237.15--55237.62 & 2.66 & 2.20(fix) & 1.5$\pm$0.7 & 0.60$\pm$0.05 & 36.0$_{-7.1}^{+9.3}$ & 6.0$\pm$0.4 & 0.79(26) \\
28 & 55237.87--55238.45 & 2.84 & 2.20(fix) & 1.7$\pm$0.6 & 0.58$\pm$0.05 & 39.3$_{-8.1}^{+10.8}$ & 5.8$\pm$0.4 & 0.96(27) \\
29 & 55238.82--55239.14 & 1.56 & 2.20(fix) & 0.5$_{-0.5}^{+0.8}$ & 0.61$\pm$0.06 & 34.3$_{-8.4}^{+12.1}$ & 5.9$\pm$0.5 & 1.32(18) \\
30 & 55239.84--55240.99 & 1.60 & 2.20(fix) & 1.8$\pm$0.9 & 0.58$_{-0.06}^{0.07}$ & 37.5$_{-10.0}^{+15.0}$ & 5.6$\pm$0.5 & 1.16(20) \\
31 & 55247.06--55247.89 & 3.22 & 2.20(fix) & 0.7$\pm$0.7 & 0.55$\pm$0.05 & 40.2$_{-8.9}^{12.0}$ & 4.7$\pm$0.3 & 0.73(23) \\
32 & 55247.95--55248.84 & 3.24 & 2.20(fix) & 1.1$_{-0.6}^{+0.3}$ & 0.55$_{-0.04}^{+0.05}$ & 42.1$_{-9.0}^{+12.1}$ & 4.6$\pm$0.3 & 0.61(24) \\
33 & 55248.91--55249.86 & 3.17 & 2.20(fix) & 1.0$\pm$0.6 & 0.57$\pm$0.05 & 36.0$_{-7.7}^{+10.4}$ & 4.2$\pm$0.3 & 0.92(24) \\
34 & 55249.92--55250.94 & 3.32 & 2.20(fix) & 0.5$\pm$0.5 & 0.55$_{-0.04}^{+0.05}$ & 39.1$_{-8.3}^{+11.1}$ & 4.2$\pm$0.3 & 0.96(21) \\
35 & 55251.00--55252.02 & 3.22 & 2.20(fix) & 0.7$\pm$0.6 & 0.56$\pm$0.05 & 36.7$_{-8.0}^{+11.0}$ & 4.2$\pm$0.3 & 0.41(23) \\
36 & 55252.08--55254.24 & 5.51 & 2.20(fix) & 0.9$\pm$0.4 & 0.53$_{-0.03}^{+0.04}$ & 41.6$_{-7.6}^{+9.7}$ & 3.7$\pm$0.2 & 0.84(28) \\
37 & 55257.09--55259.12 & 3.83 & 2.20(fix) & 1.0$\pm$0.5 & 0.54$\pm$0.05 & 36.9$_{-8.9}^{+12.5}$ & 3.1$\pm$0.2 & 0.66(23) \\
38 & 55259.52--55261.16 & 6.53 & 2.20(fix) & 0.8$\pm$0.4 & 0.52$\pm$0.04 & 41.4$_{-8.2}^{+10.8}$ & 3.1$\pm$0.2 & 0.53(26) \\
39 & 55261.49--55263.19 & 6.82 & 2.20(fix) & 0.3$\pm$0.3 & 0.55$_{-0.03}^{+0.04}$ & 34.7$_{-6.0}^{+7.6}$ & 3.3$\pm$0.2 & 0.99(27) \\
40 & 55263.44--55265.22 & 8.85 & 2.20(fix) & 0.7$\pm$0.3 & 0.51$\pm$0.03 & 40.4$_{-7.1}^{+9.0}$ & 2.9$\pm$0.1 & 0.54(29) \\
41 & 55265.35--55266.94 & 8.52 & 2.20(fix) & 0.4$\pm$0.3 & 0.50$_{-0.03}^{+0.04}$ & 42.4$_{-8.1}^{+10.6}$ & 2.8$\pm$0.1 & 1.11(25) \\
42 & 55267.00--55268.66 & 8.37 & 2.20(fix) & 0.6$\pm$0.3 & 0.50$_{-0.03}^{+0.04}$ & 42.4$_{-8.2}^{+10.6}$ & 2.6$\pm$0.1 & 0.94(27) \\
43 & 55268.72--55272.02 & 16.93 & 2.20(fix) & 0.6$\pm$0.2 & 0.49$\pm$0.03 & 43.7$_{-6.9}^{+8.5}$ & 2.3$\pm$0.1 & 0.94(36) \\
44 & 55272.28--55276.41 & 12.51 & 2.20(fix) & 0.5$\pm$0.2 & 0.46$\pm$0.03 & 51.3$_{-10.5}^{+13.9}$ & 2.0$\pm$0.1 & 0.72(29) \\
45 & 55277.23--55281.63 & 3.98 & 2.20(fix) & 0.8$\pm$0.5 & 0.47$_{-0.08}^{+0.09}$ & 39.0$_{-16.6}^{+34.2}$ & 1.4$\pm$0.2 & 0.94(14) \\
46 & 55285.44--55286.97 & 2.34 & 2.20(fix) & 2.1$_{-0.5}^{+0.4}$ & 0.31$_{-0.14}^{+0.20}$ & 118.7$_{-111.7}^{+2179.5}$ & 0.5$\pm$0.2 & 0.46(9) \\
47 & 55287.41--55288.05 & 2.42 & 2.63$_{-0.40}^{+0.47}$ & 2.0$\pm$0.4 & -- & -- & -- & 0.39(9) \\
48 & 55288.43--55291.16 & 8.98 & 2.19$_{-0.19}^{+0.21}$ & 2.0$\pm$0.2 & -- & -- & -- & 0.58(32) \\
49 & 55291.35--55292.49 & 3.41 & 1.60$_{-0.26}^{+0.27}$ & 2.7$_{-0.4}^{+0.5}$ & -- & -- & -- & 0.61(14) \\
\hline
\multicolumn{9}{@{}l@{}}{\hbox to 0pt{\parbox{100mm}{\footnotesize 
Note. 
\par\noindent
\footnotemark [$*$] In unit of 10$^{-9}$ erg s$^{-1}$ cm$^{-2}$ (2--20 keV).
}\hss}}
\end{longtable}

The MCD plus power-law model or single power-law model
gave good fits for all the MAXI/GSC data. 
Table~\ref{tab:parametertab} summarizes the obtained spectral
parameters, while figure~\ref{fig:time_vs_parameters} plots the evolution of
the model parameters.
Figure~\ref{fig:nufnuspec} shows typical response-unfolded $\nu F_{\nu}$
spectra, together with their best-fit models.

The spectra during MJD 55200--55214 (DataID = 1--10) required no MCD
component. The photon indices and fluxes was almost constant meanwhile at
$\approx$1.7 and 1.2$\times$10$^{-8}$ erg cm$^{-2}$ s$^{-1}$,
respectively. On MJD 55215.64--55216.15 (DataID = 11--12), the
spectrum dramatically softened to $\Gamma \sim$2.0. The MCD component
then appeared after MJD 55216.65 (DataID = 13) and lasted 
until MJD 55286 (DataID = 46), although the power-law flux was
dominant MJD 55216.65--55218.37 and on MJD 55286.97 in the 2-20 keV
band. The MCD flux reached to a peak of
$\sim$1$\times$10$^{-8}$erg s$^{-1}$ cm$^{-2}$ on 2010 January 23 (MJD
55219), and then decreased. The innermost temperature $T_{\rm in}$
gradually decreased from $\sim$0.7 keV to $\sim$0.4 keV toward MJD
55286.97. By contrast, the innermost radius $r_{\rm in}$ was almost
constant at $\sim$41 km (for $D=3.5$ kpc and $i=0^\circ$) then,
except for the epoch before the state transition on MJD 55218.37
(DataID = 13--15).

Based on these results, we identify that the source was likely in the
low/hard state before MJD 55214, the intermediate state over MJD
55215--55218, the high/soft state over MJD 55219--55282, and then came
back to the low/hard state after MJD 55292 through the intermediate
state between MJD 55285 and 55292. For later discussions, we
calculate the weighted average of $r_{\rm in}$ in the high/soft state to
be 41.0$\pm$2.1 km, using the observations on MJD 55230--55282 when
the contribution of the power-law component was sufficiently small.

As \citet{simpl1} pointed out the initial variation of $r_{\rm in}$ 
observed on MJD 55216--55218 in the intermediate state may be caused 
by the ignoring of Comptonized photons in the calculation of $r_{\rm in}$.
In these epochs, the fraction of the power-law component was
$>$50\% of the total X-ray flux in the 2-20 keV band, and hence
the Compton up-scattering of the disk blackbody emission can
significantly reduce the normalization of the direct MCD component
(\cite{review}). We find it difficult, however, to obtain a reliable
estimate of the innermost radius corrected for this effect from the
MAXI/GSC data alone, which strongly couples with the photon index of
the Compton scattered component. We leave detailed investigation of
the spectra in the intermediate state for future work, and
concentrate on those in the high/soft state in the following analysis.

\section{Swift Observations and Analysis}

From 2009 October 25 to 2010 July 29, the Swift/XRT (\cite{swiftxrt}) 
carried out 67 pointing observations of XTE\ J1752--223 in the Windowed Timing (WT) mode. 
Due to the Sun angle constraint, the observations
were interrupted between 2009 November 4 (MJD 55139) and 2010 February 3
(MJD 55230). We analyzed data of 17 observations taken in the high/soft
state after 2010 February 4 (MJD 55231). All the XRT spectra and light
curves were produced by the web interface (\cite{swiftukwebana})
supplied by the UK Swift Science Data Centre at the University of
Leicester. We used the XRT response matrix file for the WT mode
version 12.  All the 17 observations with an exposure time of a few
ksec each were analyzed independently.

First results of the Swift/XRT spectral analysis were already reported by
\citet{1752curran}. They employed the same MCD plus power-law model as
we used in the MAXI/GSC analyses, but derived a different result
that the innermost radius changed significantly with time.
The difference is considered to come from the fact that they left the
power-law index free and obtained a wide range of $\Gamma$ between
$\sim$0 and $\sim$3, even though this quantity is not constrained 
when the MCD component is strong.
Hence we reanalyzed the XRT spectra by fixing $\Gamma$ at 2.2, 
the same value as adopted for the MAXI/GSC spectra. 
The $N_{\rm H}$ was left free because the Swift/XRT
is sensitive down to $\sim$0.3 keV.  Resultant parameters are
summarized in table \ref{tab:parametertab_xrt}, which are also plotted
on figure~\ref{fig:time_vs_parameters}. 
The values of $r_{in}$ were thus constant, with at most 5\% scatter around 
weighted mean of $r_{in}$=41.4 $\pm$0.3 km.

\begin{longtable}{ccccccccc}
\caption{Best-fit parameters of Swift/XRT observations.}
\label{tab:parametertab_xrt}
 \hline\hline
    ObsID & MJD & Exposure & N$_{\rm H}$ & power-law & $T_{\rm in}$  & $r_{\rm in}$  & Disk  & $\chi^{2}_{\nu}$ /dof \\
  {\ } & start  & (sec) &  ($\times$10$^{22}$ cm$^{-2}$) & flux$^*$ & (keV) & (km)  & flux$^*$ & {\ }  \\    
\hline
\endhead
00031532010 & 55231.6 & 1374.3 & 0.56$\pm$0.01 & 2.34$\pm$0.23 & 0.60$\pm$0.01 & 38.4$_{-1.0}^{+1.1}$ & 5.66$\pm$0.20 & 1.68(224) \\
00031532011 & 55233.7 & 1595.0 & 0.60$\pm$0.01 & 2.40$\pm$0.19 & 0.59$\pm$0.01 & 38.0$\pm$1.0 & 5.39$\pm$0.14 & 1.29(238) \\
00031532012 & 55238.9 & 2360.7 & 0.62$\pm$0.01 & 1.26$\pm$0.11 & 0.58$\pm$0.01 & 44.3$_{-0.7}^{+0.8}$ & 6.39$\pm$0.11 & 1.50(244) \\
00031532014 & 55242.0 & 1421.4 & 0.56$\pm$0.01 & 2.41$\pm$0.19 & 0.57$\pm$0.01 & 42.2$\pm$1.1 & 5.21$\pm$0.15 & 1.64(234) \\
00031532015 & 55244.0 & 1123.8 & 0.55$\pm$0.01 & 1.07$\pm$0.17 & 0.56$\pm$0.01 & 40.9$_{-1.2}^{+1.3}$ & 4.31$_{-0.16}^{+0.17}$ & 1.22(191) \\
00031532016 & 55246.2 & 1411.8 & 0.59$\pm$0.01 & 2.80$\pm$0.18 & 0.57$\pm$0.01 & 41.7$\pm$1.1 & 4.79$_{-0.12}^{+0.13}$ & 1.56(241) \\
00031532018 & 55248.0 &  835.6 & 0.63$\pm$0.01 & 1.29$\pm$0.16 & 0.55$\pm$0.01 & 45.3$\pm$1.4 & 5.01$_{-0.13}^{+0.14}$ & 0.95(202) \\
00031532019 & 55250.1 & 1144.3 & 0.61$\pm$0.01 & 0.47$\pm$0.12 & 0.54$\pm$0.01 & 41.7$\pm$1.4 & 3.96$_{-0.14}^{+0.16}$ & 1.22(177) \\
00031532020 & 55251.3 & 3026.5 & 0.61$\pm$0.01 & 0.87$\pm$0.07 & 0.55$\pm$0.01 & 41.9$\pm$0.7 & 4.07$\pm$0.07 & 1.62(238) \\
00031532021 & 55252.6 & 1524.0 & 0.57$\pm$0.01 & 0.65$\pm$0.09 & 0.53$\pm$0.01 & 43.4$\pm$1.1 & 3.80$\pm$0.06 & 1.50(193) \\
00031640001 & 55257.6 &  890.9 & 0.60$\pm$0.01 & 0.69$\pm$0.11 & 0.52$\pm$0.01 & 42.5$_{-1.5}^{+1.6}$ & 3.05$\pm$0.08 & 1.22(174) \\
00031640002 & 55260.7 &  909.9 & 0.61$\pm$0.01 & 0.25$\pm$0.05 & 0.52$\pm$0.01 & 36.9$_{-1.0}^{+1.1}$ & 2.23$_{-0.08}^{+0.09}$ & 1.34(180) \\
00031640003 & 55266.8 & 1209.8 & 0.61$\pm$0.01 & 0.17$\pm$0.06 & 0.50$\pm$0.01 & 43.4$_{-1.4}^{+1.5}$ & 2.38$\pm$0.09 & 1.25(161) \\
00031640004 & 55269.8 & 1184.1 & 0.67$\pm$0.01 & 0.32$\pm$0.04 & 0.50$\pm$0.01 & 39.3$_{-0.9}^{+1.0}$ & 2.00$\pm$0.05 & 1.61(197) \\
00031640005 & 55271.8 & 1214.6 & 0.66$\pm$0.01 & 0.33$\pm$0.04 & 0.48$\pm$0.01 & 43.3$\pm$1.3 & 1.75$\pm$0.06 & 1.41(181) \\
00031640006 & 55276.9 &  977.8 & 0.57$\pm$0.01 & 0.34$\pm$0.04 & 0.48$\pm$0.01 & 41.5$_{-1.1}^{+1.2}$ & 1.67$\pm$0.04 & 1.34(177) \\
00031640007 & 55279.8 &  128.3 & 0.54$_{-0.02}^{+0.03}$ & $\pm$0.12 & 0.47$\pm$0.01 & 35.8$_{-2.8}^{+3.1}$ & 0.99$_{-0.10}^{+0.11}$ & 1.00(108) \\
  \hline
\multicolumn{9}{@{}l@{}}{\hbox to 0pt{\parbox{100mm}{\footnotesize 
Notes. 
\par\noindent
\footnotemark []Photon indices was fixed to 2.20.
\par\noindent
\footnotemark [$*$] In unit of 10$^{-9}$ erg s$^{-1}$ cm$^{-2}$ (2--20 keV).
}\hss}}
\end{longtable}

\section{Suzaku Observations and Results}

\subsection{Observation}

We also observed XTE\ J1752--223 in the high/soft state with Suzaku
(\cite{suzaku}) as a
Target-of-Opportunity (ToO) from 2010 February 24 (MJD 55251) 04:58:00 to
February 25 04:27:24 (ObsID: 904005010). 
Suzaku carries 4 X-ray telescopes (XRT, \cite{suzaku_xrt}), each with a
focal-plane X-ray CCD imager called XIS (X-ray Imaging Spectrometer;
\cite{suzaku_xis}) covering the 0.2--12 keV energy band. 
Since XIS2 has not been available, the two remaining front-illuminated
(FI) CCDs (XIS0 and XIS3) were operated for this observation together
with the back-illuminated (BI) CCD (XIS1). The three XIS cameras were operated
with burst and window options. The burst time, window size, and editing
mode were ``0.3 sec, 1/4 window, and 2$\times$2'' for XIS0 and XIS3, and
``0.1 sec, normal window, and 3$\times$3'' for XIS1, respectively.
The Hard X-ray Detector (HXD; \cite{suzaku_hxd_takahashi};
\cite{suzaku_hxd_kokubun}), covering the 10--70 keV energy band with
Si PIN photo-diodes (HXD-PIN), and the 50--600 keV range with GSO
scintillation counters (HXD-GSO), was operated in the normal mode. The
source was observed at the HXD nominal position.

\subsection{Data Reduction}

The data reduction and analyses were performed using Suzaku
{\footnotesize FTOOLS} in {\footnotesize HEASOFT} version 6.10
provided by the NASA/GSFC Suzaku GOF.  All the XIS and HXD data were
reprocessed by the Suzaku pipeline processing version 2.4.12.27, with
the latest calibration database ({\footnotesize CALDBVER})
hxd20090902, xis20090925, and xrt20080709.

The image degradation due to ``thermal wobbling'' of the satellite
pointing was corrected by using the attitude file updated with 
{\footnotesize AEATTCOR2}. We screened the XIS data under the standard
criteria, and discarded events during time intervals when the telemetry
of the XIS was saturated. Resultant net exposure was 10.6 ksec and 449
sec for FIs and BI, respectively.
The BI exposure was short, because 0.1 sec burst time with normal window
make the exposure 0.1 sec in 8 sec period and the exposure is one
twelfth of each FI's exposure.
Furthermore, the fraction of out-of-time events (\cite{suzaku_xis}) is 
more significant in the BI data, which would introduce 
systematic uncertainties in the spectral analysis.
Hence, we did not use the XIS1 data in this paper.
The average 0.5--10 keV count rate was
$\sim$400 cts s$^{-1}$ and $\sim$700 cts s$^{-1}$ with FIs and BI,
respectively. 
We extracted the XIS0 and 3 events
from a rectangle region of \timeform{8'.6}$\times$\timeform{4'.5}
tracing the 1/4 window area centered on the image peak,
and the background from another region avoiding contaminating point sources.
 The source was so bright that the XIS
suffered from photon pileup at the image center. Using the software
of {\footnotesize AEPILEUPCHECKUP} (\cite{suzaku_pileup} in
preparation), which automatically detects the extent of pileup utilizing
radial surface brightness profiles and other sets of information, we
excluded a circular within  \timeform{1'.5} of the image peak for
XIS0 and 3, to extract the events whose pileup fraction is below 1\%.
We combined the spectra and response files of the two FI cameras to
increase the photon statistics in the spectra.
To account for possible uncertainties in the instrument calibration
(e.g. \cite{cygx1_makishima}), systematic errors of 1\% was assigned to
each energy bin of the XIS spectra. 

We obtained the cleaned HXD events with the standard criteria.
After the dead-time correction, the net exposures of PIN and GSO were
21.2 and 27.0 ksec, respectively.
The dead-time fraction was 18.2 and 7.5 \% for PIN and GSO,
respectively.
The cosmic X-ray background
(CXB) was modeled assuming an exponentially cutoff power-law model
(\cite{boldt_cxb}). 
The non-X-ray background (NXB) model is provided
by the HXD team (\cite{hxd_bgd}). We used the model with
{\footnotesize METHOD}=``{\footnotesize LCFITDT}(bgd\_d)'', and the
version of {\footnotesize METHODV}=``2.0ver0804'' and
``2.4ver0912-64'' for the data of PIN and GSO, respectively.  
After subtracting the modeled NXB and CXB, the source signals were
significantly detected by PIN up to 50 keV, above the systematic
uncertainty on the NXB model ($\sim$3\%). Since the GSO signals were
comparable to the systematic errors of the NXB model there in
($\sim$1\%), we consider the source detection with GSO insignificant.

\subsection{Spectral Analysis}
\label{suzakuresults}

\begin{figure}[!ht]
 \begin{center}
  \FigureFile(75mm,50mm){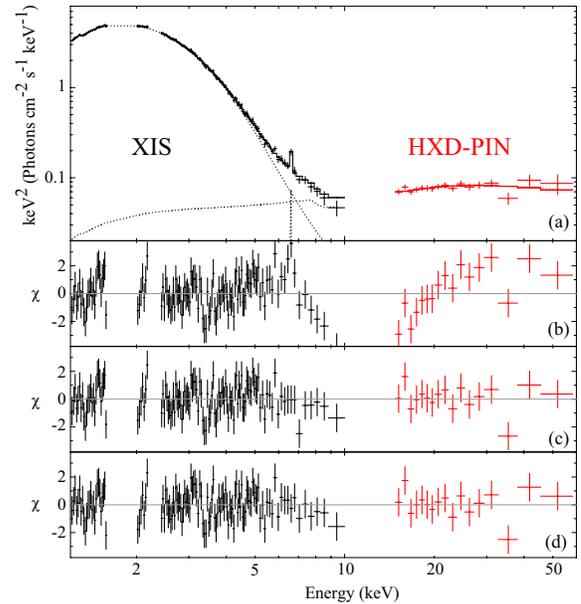}
 \end{center}
   \caption{ (a) Suzaku response-unfolded $\nu F\nu$ spectrum of XTE J1752--223 from the best-fit
  for {\em wabs*(compPS+pexriv+gaus)}, and the residuals between the data and best-fit models for
  (b) {\em wabs*(diskbb+powerlaw)},
  (c) {\em wabs*smedge*(diskbb+powerlaw+gaus)},
  (e) {\em wabs*(compPS+pexriv+gaus)}.
  }
\label{suzaku_specfit}
\end{figure}

As shown in figure~\ref{suzaku_specfit}(a), we used an
energy range of the 1.2--10 keV for the XIS spectrum and 15--50 keV for
the HXD-PIN spectrum, where the calibrations of the energy responses were
well established.  
The energy bands of 1.6--2.0 keV and 2.2--2.4 keV were excluded to
avoid large systematic uncertainties of the effective area near the
Silicon K edge (1.74 keV and 1.84 for K$\alpha$ and K$\beta$,
respectively) and gold M edge (2.29 keV), respectively.  
In the combined fit to the XIS and HXD-PIN spectra, we employed a cross
normalization factor of 1.18 between XIS and PIN (SUZAKU-MEMO-2008-06).

We firstly fitted the spectra with the MCD plus power-law model
modified by interstellar absorption ({\em wabs} model), the same one
as adopted to fit the MAXI/GSC and Swift/XRT spectra in the previous
sections. The residuals in units of $\chi$ are shown in the
figure~\ref{suzaku_specfit}(b). The fit was found to be unacceptable
($\chi^2/dof=230.1/136$), leaving an emission-line feature at
$\approx$6.5 keV, and significant broad edge-like residuals above
$\sim$7 keV. These features may be explained away by several different
approaches; e.g., by invoking a broad Fe-K line concept 
\citep{1752_ironline} or
applying so-called smeared-edge model (e.g., \cite{smedge}; 
\cite{dotani_mcd}) to account for the edge structure.
The latter approach with {\em wabs*smedge*(diskbb+powerlaw+gaus)} model  
gave an acceptable fit ($\chi^2/dof=141.6/133$; fig~\ref{suzaku_specfit}({c}))
with a maximum optical depth of $5.50_{-0.75}^{+1.4}$ for fixed edge energy and
width of 7.11 keV and 10 keV, respectively. A narrow emission line was
found at $E_{\rm c}$=6.59$_{-0.07}^{+0.08}$ keV with a 1$\sigma$ line width of 10
eV (fixed) and an equivalent width of $EW$=63$_{-41}^{+42}$ eV. These
approaches are empirical and may be degenerate, however.

As an alternative possibility, we resorted to the idea of
\citet{Kolehmainen} that the disk emission is broader than the
simplest MCD model, which could lead to artificial detection of broad
iron-K features. Assuming that weak Comptonization of the MCD emission
took place, we replaced the {\em diskbb} model with a {\em compPS}
model (\cite{compps}), which computes Compton scattering by hot corona using
exact numerical solution radiative transfer equation. We chose
``slab'' (geom=1) geometry and a MCD spectrum as the seed photons. We
left the optical depth of the corona $\tau$ as a free parameter but
fixed the electron temperature at $T_{\rm e}=10$ keV and covering
fraction at ${\rm cov_{frac}}=1$ to avoid strong coupling between
these parameters; the choice of $T_{\rm e}$ did not affect our results
on the disk parameters over the statistical errors. The presence of a
fluorescence iron-K line at $\approx$6.6 keV indicates that a
reflection continuum must be included in the model as well. Hence, we
adopted {\em pexriv} model (\cite{pexriv}) for the power-law component, which was
considered to be the main source of irradiation onto the disk in the
high/soft state. The model thus became {\em wabs*(compPS+pexriv+gaus)}
in the XSPEC terminology. For the {\em pexriv} component, we assumed
no high energy cutoff, and fixed the solid angle of the reflector at
$\Omega=2\pi$, inclination at $25^\circ$, and ionization parameter
at $\xi_{\rm i}=1000$ erg cm s$^{-1}$, which consistently explain both iron-K line intensity
and center energy. The fit was found to be acceptable with $\chi^2/dof
= 129.9/132$; as shown in the figure~\ref{suzaku_specfit}(d), the
broad edge-like residuals mostly disappeared. The model gave an
absorbed 2--20 keV flux of (3.76$\pm$0.02)$\times$10$^{-9}$\ {\rm erg}\ s$^{-1}$cm$^{-2}$. 
We obtained $\tau=0.41\pm0.05$ and $\Gamma=2.13\pm0.01$ 
in the {\em pexriv} component, a reasonable value in
the high/soft state (e.g., \cite{zdziarski_review}). More detailed
analysis, more advanced disk emission modeling, and relativistic Fe-K
lines, is beyond the scope of this paper, and will be reported
elsewhere.

The best-fit parameters of these models are listed in
table~\ref{suzaku_bestfit}. We note that by replacing the {\em diskbb}
model with the {\em compPS} model, the best-fit value of $r_{\rm in}$
increased by 13\%. This is because a spectrum hardened by
Comptonization was assumed in the {\em compPS} model, an intrinsic
temperature $T_{\rm in}$ became lower, leading to an increase in
$r_{\rm in}$. While the difference should be regarded as a systematic
modeling uncertainty, we confirm that it does not affect our
conclusion.
The photon index of the hard tail ($\Gamma=2.13\pm0.01$) is
smaller than the result obtained from the same Suzaku data by \citet{1752_ironline}
($\Gamma=2.54_{-0.11}^{+0.02}$), who did not include the 15-20 keV and 45-50 keV
bands in fitting the HXD/PIN spectrum.

\begin{table*}[!ht]
 \caption{Best-fit parameters for the Suzaku observation.}
 \label{suzaku_bestfit}
 \begin{center}
  \begin{tabular}{ c c c c c c }
   \Hline
   \multicolumn{6}{l}{Model} \\
   \Hline
   {\em wabs} & {\em diskbb}\ /\ {\em compPS}$^{\dagger}$ & {\em powerlaw}\ /\ {\em pexriv}$^{\|}$ & {\em gaus}$^{**}$ & {\em smedge}$^{\dagger\dagger}$  & \ \\
   \hline
   $N_{\rm H}^{*}$ & $T_{\rm in}$ (keV) &  $\Gamma$ &  $E_{\rm c}$ (keV) & $\tau_{\rm max}^{\ddagger\ddagger}$  & $\chi^2$ / d.o.f. \\
   \  & $r_{\rm in}^{\ddagger}$ (km)  &  norm$^{\#}$ & $EW$ (eV) & \  \\
   \  & $\tau^{\S}$  & \ & \   & \ \\
   \Hline
   \multicolumn{5}{l}{ {\em wabs*(diskbb+powerlaw)} } \\
   \hline
   0.62$\pm0.02$ & 0.53$\pm0.01$ & 2.29$\pm$0.06  & {...}  & {...} & 230.1/136 \\
   {...}   & 42.9$\pm0.9$ & 0.16$\pm$0.03  & {...} & {...}  & {...}   \\
   {...}   & {...}   & {...}   & {...} & {...}  & {...}   \\
   \Hline
   \multicolumn{6}{l}{ {\em wabs*smedge*(diskbb+powerlaw+gaus)} } \\
   \hline
   0.65$\pm0.02$ & 0.526$\pm0.003$ & 2.22$\pm$0.05  & 6.59$_{-0.07}^{+0.08}$  & $5.50_{-0.75}^{+1.4}$ & 141.80/133 \\
   {...}   & 44.0$\pm1.0$ & 0.16$\pm$0.02  & 63$_{-41}^{+42}$  & {...} & {...}   \\
   {...}   & {...}   & {...}   & {...}  & {...} & {...}   \\
   \Hline
   \multicolumn{6}{l}{ {\em wabs*(compPS+pexriv+gaus)} } \\
   \hline
   0.67$\pm0.02$   & 0.507$_{-0.006}^{+0.005}$  & 2.13$\pm0.01$ & 6.60$_{-0.07}^{+0.08}$ & {...} & 131.4/133 \\
   {...}   & 47.2$\pm0.8$  & 0.05$_{-0.04}^{+0.08}$ &  98$\pm43$ & {...} & {...}   \\
   {...}   & 0.41$\pm0.05$  & {...}   &  {...} & {...} & {...}   \\
   \hline
\multicolumn{6}{@{}l@{}}{\hbox to 0pt{\parbox{100mm}{\footnotesize 
Notes. 
\par\noindent
\footnotemark [$*$] In unit of 10$^{22}$ cm$^{-2}$.
\par\noindent
\footnotemark [$\dagger$] $T_{\rm e}$ and cov$_{\rm frac}$ are fixed at 10 keV and at 1, respectively. No reflection from the {\em compPS} component itself is considered.
\par\noindent
\footnotemark [$\ddagger$]$D$=3.5 kpc and $i=0^\circ$ are assumed.
\par\noindent
\footnotemark [$\S$] Optical depth of the corona.
\par\noindent
\footnotemark [$\|$] Fixed at $\Omega=2\pi$, $i=25^\circ$ and $\xi_{\rm i}=1000$ erg cm/s. No high energy cutoff is set in the incident power law continuum.
\par\noindent
\footnotemark [$\#$] A normalization in unit of photons cm$^{-2}$ s$^{-1}$ at 1 keV.
\par\noindent
\footnotemark [$\**$] 1 $\sigma$ line width is fixed at 10 eV.
\par\noindent
\footnotemark [$\dagger\dagger$] The edge energy and width, photo-electric cross-section are fixed at 7.11 keV, 10 keV and -2.67, respectively.
\par\noindent
\footnotemark [$\ddagger\ddagger$] The maximum absorption factor at threshold.

}\hss}}
\end{tabular}
\end{center}
\end{table*}

\section{Discussion}

We have analyzed the MAXI/GSC, the Swift/XRT, and
Suzaku data of XTE\ J1752--223 in the high/soft state from MJD
55218.62 to 55281.63. The overall continuum spectra were reproduced by
the MCD plus power-law model with an interstellar absorption. The
innermost temperature $T_{\rm in}$ decreased from $\sim$0.7 keV to
$\sim$0.4 keV, while the innermost radius $r_{\rm in}$ remained
constant. By assuming $D=3.5$ kpc and $i=0^\circ$, the values of
$r_{\rm in}$ derived from the MAXI/GSC, Swift/XRT, and Suzaku data are
41.0$\pm$2.1 km, 41.4$\pm$0.3 km, and 42.9$\pm0.9$, respectively.  The
results from the three instruments are mutually consistent with each
other within the statistical errors. For the following discussion, we
employ 41.0$\pm$2.1 km for $r_{\rm in}$ as determined from the
MAXI/GSC, although we also discuss the case when the Suzaku result
with the {\em compPS} model ($r_{\rm in}$=47.2$\pm0.8$ km) is adopted to
take into account the possible systematic uncertainty.

\begin{figure}
  \begin{center}
    \FigureFile(80mm,80mm){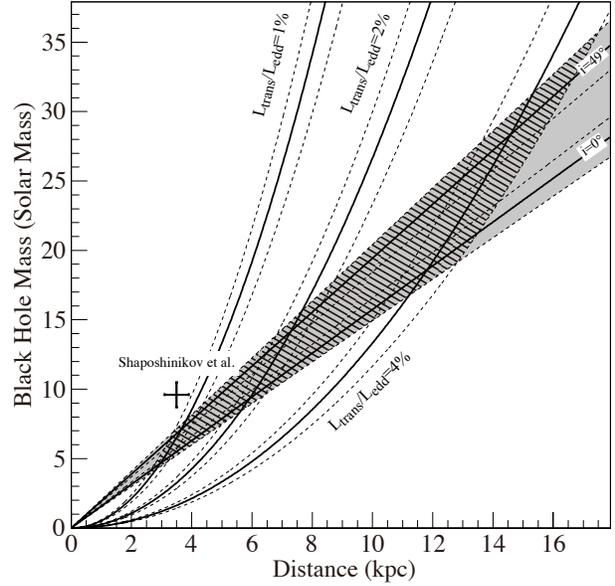}
  \end{center}
  \caption{Observational constraints on the mass and distance of XTE\,J1752--223.
   The shadowed region is specified by the innermost radius derived from the MAXI/GSC spectra   
  in combination with the inclination angle ${\it i}$$<$49$^{\circ}$ from the radio observation (\cite{accurate_pos_vlbi}).
  Each solid line with two dashed lines indicates the best fit parameter and 90\% confidence ranges of 
  flux and $r_{\rm in}$.
  The hatched region is derived assuming the empirical relation
  that the soft-to-hard transition occurs at 1$\sim$4 \%$L_{edd}$.}
  \label{fig:distancemass}
\end{figure}

Since the value of $r_{in}$ thus estimated is significantly larger
than those found in luminous low-mass X-ray binaries (~10 km;
\cite{mcdmodel}), the black hole interpretation of XTE\,J1752--223
(Paper I; \cite{1752_xte_lhs}) is considerably reinforced.
Furthermore, its constancy allows us to identify it with the ISCO in
the high/soft state.  The stable $r_{\rm in}$ at 41.0 km is supposed
to reach the ISCO in the high/soft state. We note that this $r_{\rm
in}$ is an ``apparent'' innermost radius, and the ``realistic''
innermost radius ($R_{\rm in}$) should be estimated as $R_{\rm in}$=
$\xi\kappa^2r_{\rm in}$ where 
the spectral hardening factor$\kappa$ is 1.7 (\cite{mcd_hardning}) and correction factor for the boundary condition
 $\xi$ is 0.412 (\cite{true_Rin}).
The value of $\kappa$ has been confirmed in recent work on disk models
(e.g., \cite{lmcx3_kubota}; \cite{angular_mom_done}).
When the central object is assumed to be a non-spinning black hole (i.e. Schwarzschild
black hole), $R_{\rm in}$ should equal to $6 R_{\rm g}$.  Then, the
black-hole mass is estimated as
\begin {eqnarray}
\label{eq-mass}
M = \frac{c^2R_{\rm in}}{6G} = 5.51\pm0.28\left(\frac{D}{3.5 {\rm kpc}}\right) (\cos{\it i})^{-\frac{1}{2}}  M_{\odot} .
\end {eqnarray}
Using the inclination angle $i<$49$^{\circ}$ 
obtained from radio observations (\cite{accurate_pos_vlbi}), 
the black hole mass and distance 
are constrained in the shadowed area shown in figure~\ref{fig:distancemass}.
If we assume the distance to be 3.5 {\rm kpc}, the mass will be
5.2--7.1 $M_{\odot}$, and if $D$=10 {\rm kpc}, 15.0--20.4 $M_{\odot}$.

To validate our method to estimate the black hole
mass by employing the MCD model which is a simplified approximation
of a true disk spectrum, we also performed a spectral fit to the
Suzaku spectra using the {\em kerrbb} model (\cite{kerrbb}). This model
calculates the X-ray spectrum of a relativistic accretion disk around
a rotating black hole by taking into account the innermost boundary
condition and the effects of self-irradiation of the disk. Here we
fixed the distance at $D$=3.5 kpc, spin parameter at $a$=0, and color
correction factor at $\kappa$=1.7 for consistency with the previous
discussion. The fit with the {\em wabs*(kerrbb+pexriv)} model was not
acceptable ($\chi^2/dof=215.5/135$), leaving similar residuals seen
in figure\ref{suzaku_specfit}(b). Thus, we employed the {\em simpl}
model (\cite{simpl1}) with a steep photon index fixed at 7.4 to
approximately represent weak Comptonization of the disk emission; the
model became {\em wabs*(simpl*kerrbb+pexriv+gaus)}, which was found to
give an acceptable fit ($\chi^2/dof=129.5/134$). The black hole mass
derived for $i$=0$^{\circ}$ and $i=$49$^{\circ}$ is
$4.98_{-0.25}^{+0.28}$ M$_{\odot}$ and 8.43$_{-0.42}^{+0.44}$ M$_{\odot}$,
respectively, which differs by $\sim$20\% 
from the mass presented in equation (1). 

From past observations of BHCs and neutron stars \citet{ltrans} pointed out
that the state transition from the high/soft state to the
low/hard state occur at 1-4\% (centered at 2\%) of the Eddington
luminosity
\footnote{$L_{\rm edd}$ = 1.5$\times$10$^{38}$
$\frac{M}{M_{\odot}}$erg s$^{-1}$ for the solar abundances}.
Judging from the change of the photon index, XTE\
J1752--223 went back to the low/hard state around MJD 55292 (DataID
49). The bolometric flux at the transition to be $F_{\rm
trans}$=(6.68$\pm$0.83)$\times$10$^{-9}$ erg cm$^{-2}$ s$^{-1}$
assuming a cutoff power-law is continuum with $\Gamma=1.8$ 
and a cutoff energy of 200 keV. Using the relation
0.01$<L_{\rm trans}$/$L_{\rm edd}<$0.04 and $L_{\rm trans}=4\pi D^2 F_{\rm bol}$, 
we can further constrain the distance and black hole mass of
XTE\ J1752--223 as illustrated with the hatched region in figure~\ref{fig:distancemass}.
Thus we can obtain $D$=2.5--18 {\rm kpc} and $M$=3.7--36 $M_{\odot}$.
When we instead use the Suzaku results, 
$D$=3.0--20 {\rm kpc} and $M$=5.3--44 $M_{\odot}$ (based on the {\em
wabs*(compPS+pexriv+gaus)} model) or 
$D$=2.3--22 {\rm kpc} and $M$=3.1--55 $M_{\odot}$ (based on the {\em
wabs*(simpl*kerrbb+pexriv+gaus)}) are derived.

Since we obtain $M>$3$M_{\odot}$, our method of
estimating $M$ assuming a black hole is self consistent. The
conclusion that the compact object in XTE\ J1752--223 is a black hole
is robust against the assumptions that the rotational axis
of the accretion disk and jet axis is exactly aligned and that the
black hole is non-spinning, since the extreme conditions $i=0^\circ$ and
$a$=0 yield the smallest mass estimate. 
The values of $M$=9.6$\pm$0.9 $M_{\odot}$ and $D$=3.5$\pm$0.4 {\rm
kpc} estimated by \citet{1752shapos} are outside the hatched
region. If the black hole in XTE J1752--223 is spinning and/or the disk has a
larger inclination than 49$^\circ$, then the discrepancy could be solved.

\section{Conclusions}

Using data obtained by the MAXI/GSC, the Swift/XRT,
and Suzaku, we have performed X-ray spectral analysis of the black
hole candidate XTE\,J1752--223 in the high/soft state. As commonly
seen in BHCs, the innermost radius remained constant in this state
during the continuous observation with the MAXI/GSC. The results using
the MCD plus power-law model were consistent between the three
observatories. Detailed spectral modeling using Suzaku data suggests a
possibility that the MCD emission may be slightly Comptonized, which
could explain apparently broad iron-K features. Assuming that the
innermost radius of the disk corresponds to the ISCO and the black
hole is non-spinning, we estimated the mass of the central object as
5.51$\pm$0.28 $M_{\odot}$ $\left(\frac{D}{3.5 {\rm kpc}}\right)
(\cos{\it i})^{-\frac{1}{2}}$ with corrections for the stress-free 
inner boundary condition and color hardening factor of 1.7.
Furthermore, using the observational
results that the inclination angle {\it i} is smaller than
49$^{\circ}$ from radio observations and the ``soft-to-hard''
transition occurs at 1--4 \% Eddington luminosity, the distance and
black hole mass of XTE\,J1752--223 were constrained as 3.1--55
$M_{\odot}$ and 2.3--22 {\rm kpc}, respectively, based on an 
analysis of the Suzaku spectra with a relativistic accretion disk model.
Thus we can conclude that XTE\ J1752--223 is likely a black hole. \\

This research has made use of MAXI data provided by RIKEN, JAXA and
the MAXI team. Suzaku data is provided by a collaborative mission
between the space agencies of Japan (JAXA) and the USA (NASA). We also
thank the Swift team for their observation. This work made use of
data supplied by the UK Swift Science Data Centre at the University of
Leicester. This research was partially supported by the Ministry of
Education, Culture, Sports, Science and Technology (MEXT),
Grant-in-Aid No.19047001, 20041008, 20540230, 20244015 , 20540237,
21340043, 21740140, 22740120.
One of the authors (S. N.) is grateful to a grant from the Hayakawa
Satio Fund awarded by the Astronomical Society of Japan.


\clearpage

\appendix

\renewcommand{\thechapter}{A}
\addcontentsline{toc}{chapter}{Appendix}
\setcounter{figure}{0}
\setcounter{table}{0}
\renewcommand{\thetable}{A-\arabic{table}}
\renewcommand{\thefigure}{A-\arabic{figure}}
\section{MAXI/GSC spectral calibration using the Crab Nebula}

We here show the status of the GSC energy response calibration with
Crab-nebula data to confirm the validity of the spectral analysis.  We
used event data with the process version 0.3b, which is screened with
the operation high voltage of a nominal 1650 V and the anode number
of \#0, \#3, \#4, and \#5 whose positional response had been well
established. The event selection is same as that employed in the first
performance verification in \citet{gscsugizaki}.

We here screened the event data with more severe conditions to verify
the response calibration with a better accuracy. We selected events
taken only during such a good scan transit that the source incident
angle $\phi_{\rm col}$ is $< 36^\circ$ and the area for both the
source and the background are completely covered.

The calibration of the energy response was performed using the
screened background and Crab Nebula data in the following steps; The
energy-PHA relations, which depend on the detector position along the
anode wires, are corrected by a gain factor for each output amplifier.
The parameter was calibrated using fluorescent lines in the background
spectrum from Ti (4.51 keV) and Cu (8.04 keV) as well as the
calibration source $^{55}$Fe (5.895 keV). We verified that the gain
factor obtained with a 0.1 \% precision successfully reproduce gain
position dependence within the discrepancy of 1\% in RMS throughout
the whole detector area.  The threshold and the resolution of the
Lower-Discriminator (LD) of each amplifier were then calibrated using
the Crab Nebula spectra.  The LD parameters were tuned so that the
results of the spectral fit to an absorbed power-law model agrees with
those of the canonical values, photon index $\Gamma=2.1$ and hydrogen
absorption column density $N_{\rm H}=0.35\times 10^{22}$ cm$^{-2}$
(i.e. \cite{toorseward}, \cite{standardcrab}).

\begin{figure}[!hb]
\begin{center}
\FigureFile(80mm,80mm){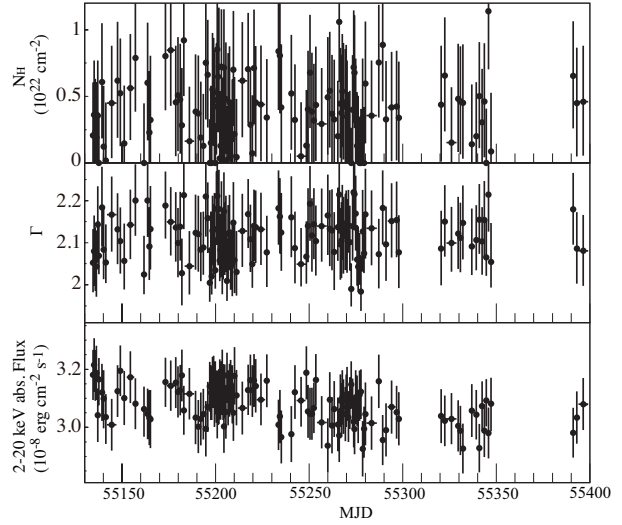}
\end{center}
\caption{
Best-fit values and statistical errors of absorbed power-law model to GSC Crab Nebula spectra
against the observation time;
(top) hydrogen column density, 
(middle) photon index,
(bottom) 2--20 keV absorbed flux.
}
\label{fig:crabana}
\end{figure}

Figure \ref{fig:crabana} shows the best-fit parameters to GSC
Crab Nebula spectra. The data of multiple scan transits 
whose total exposure of $>$3.2 cm$^2$ ksec were grouped so that each 
spectrum has enough photon statistics to constrain
parameters of a power-law model. Table \ref{table:crabresult}
summarizes the results. The derived parameters agree with the
canonical values within the statistical errors.

\begin{table}[!ht]
\caption{
Best-fit values and variations of absorbed power-law model to GSC Crab Nebula spectra
}
\label{table:crabresult}
\begin{center}
\begin{tabular}{lccc} 
  \Hline
Parameter   & Canonical & Best fit & rms$*$ \\
\hline
  $N_{\rm H}$ ($\times10^{22}$ cm$^{-2}$) & 0.35  & 0.39 & 0.25 \\
  $\Gamma$      & 2.10 & 2.11 & 0.05   \\
  $\Gamma^{\ddagger}$      & \  & 2.11 & 0.03   \\
 Flux$_{\rm 2-20 keV}$ $\dagger$     & 3.0$\pm$10\% & 3.08 & 0.06 \\
\hline
\multicolumn{4}{@{}l@{}}{\hbox to 0pt{\parbox{100mm}{\footnotesize 
Notes. 
\par\noindent
\footnotemark [$*$] Root mean square of the best-fit value
\par\noindent
\footnotemark [$\dagger$] In unit of 10$^{-8}$ erg s$^{-1}$cm$^{-2}$
\par\noindent
\footnotemark [$\ddagger$] $N_{\rm H}$ is fixed at 0.35$\times10^{22}$ cm$^{-2}$
}\hss}}

\end{tabular}
\end{center}
\end{table}

\end{document}